\documentclass[%
 reprint,
 amsmath,amssymb,
 aps,
]{revtex4-2}

\usepackage{placeins}
\usepackage{float}

\usepackage{graphicx}
\usepackage{dcolumn}
\usepackage{bm}

\usepackage[colorlinks,
            linkcolor=blue,
            citecolor=blue,
            urlcolor=blue]{hyperref}

\begin{document}

\preprint{APS/123-QED}

\title{Cosmological Impacts of Black Hole Mergers:\\ No Relief in Sight for the Hubble Tension}

\author{Zachary J. Hoelscher}
\affiliation{Department of Physics and Astronomy, Vanderbilt University, Nashville, TN 37235, USA}
\email{zachary.j.hoelscher@vanderbilt.edu} 

\author{Thomas W. Kephart}
\affiliation{Department of Physics and Astronomy, Vanderbilt University, Nashville, TN 37235, USA}
\email{thomas.w.kephart@vanderbilt.edu}  

\author{Kelly Holley-Bockelmann}
\affiliation{Department of Physics and Astronomy, Vanderbilt University, Nashville, TN 37235, USA}
\affiliation{Department of Life and Physical Sciences, Fisk University, Nashville, TN 37208, USA}
\email{k.holley@vanderbilt.edu} 

\date{\today}

\begin{abstract}
The values of the Hubble constant inferred from local measurements and the cosmic microwave background (CMB) exhibit an approximately 5$\sigma$ tension. Some have suggested this tension is alleviated if  matter is converted to dark radiation via dark matter decay. As it is not clear that dark matter decays, we instead examine the effects of converting matter to gravitational radiation via black hole mergers. We consider mergers of supermassive black holes (SMBHs), mergers of stellar-mass black holes, and the formation of SMBHs from mergers of smaller black holes. We find that these processes cannot alleviate the tension, as an unrealistically large merger rate, or an overproduction of SMBHs is required. We also consider whether one can use the Integrated Sachs-Wolfe effect to constrain mechanisms that form SMBHs from mergers of smaller black holes. We find that this is also too small to be viable.
\end{abstract}

\maketitle


\section{\label{sec:intro} Introduction}

The Hubble constant ($\rm{H_0}$) dictates how the redshift of galaxies increases with their distance from Earth, and is closely related to the expansion rate of the Universe, first studied by Edwin Hubble \cite{Hubble}. $\rm{H_0}$ can be determined through either a local measurement or an early-time measurement, and these produce different values. The local measurement involves using Type-IA supernovae as a tool to measure distances to faraway galaxies, where the distance and the redshift can be used to obtain $\rm{H_0}$. This method is calibrated using a distance ladder of Cepheid variables and parallax measurements for nearby Cepheids \cite{ModernAstrophysicsBook, Maguire2017, Local_H0}. The early-time measurement is instead based on the power spectrum of the cosmic microwave background (CMB), where the Hubble constant is tied to the location of the first peak in the power spectrum \cite{Early_Dark_Energy}. The CMB formed at around $z = 1090$, when the Universe cooled enough for electrons and protons to combine to form atoms, allowing photons to travel freely \citep{CMB_Review}. These photons have now redshifted into the microwave band. 

The CMB power spectrum implies (under the assumption of $\Lambda$CDM) a smaller value of $\rm{H_0}$ than the locally-measured value, with the two values in a $5\sigma$ tension \cite{Local_H0}\cite{Planck_H0}. Previous studies have attempted to explain this discrepancy by assuming the conversion of matter to dark radiation via dark matter decays \cite{DM_Decay}\cite{Pandey_2020}. While there is a wealth of evidence to suggest that dark matter exists \cite{Rubin}\cite{DMReview}, it is not known that it decays in this way. It is, however, known that black hole mergers convert a fraction of the mass of the binary to gravitational radiation \cite{MassLossGW}, which can be considered to be a form of dark radiation. We thus examine whether the Hubble tension can be relieved through mergers of supermassive black holes (SMBHs), stellar-mass black holes, or through forming SMBHs through mergers of smaller black holes. Others have studied the effects from conversion of matter to gravitational radiation through mergers of primordial black holes \cite{PBH}, though they did not consider mergers of more massive black holes, such as those that are located near the centers of galaxies \cite{SMBH_Formation}. Our work is complementary to past work on radiation released during structure formation \citep{StructureFormation}, and supports the conclusions therein.

We find that SMBH or stellar-mass mergers are not a viable solution for alleviating the tension, as an unreasonably large merger rate would be required. We also find that SMBH formation is not a viable pathway, as even in an extreme scenario of forming SMBHs from mergers of $10^{18}$ kg primordial black holes (PBHs), one would have to overproduce SMBHs by three to four orders of magnitude to alleviate the tension. 

The formation mechanism for SMBHs is still an outstanding mystery, though two possible scenarios involve direct collapse or mergers of smaller black holes \cite{SMBH_Formation}. Since it is known that decaying dark matter scenarios suffer from CMB constraints due to their effect on the low multipole moment region of the CMB via the Integrated Sachs-Wolfe (ISW) effect \cite{DM_Decay_Constrained}, we also consider whether one could use the CMB to constrain scenarios for SMBH formation from mergers of many smaller black holes. We find that the effect on the CMB is too small for this to be viable. 

\section{\label{sec:methods} Methods}

\subsection{\label{sec:BHMergers}Black Hole Mergers and SMBH Formation}

The code used in the following analysis is publicly available~\citep{Hoelscher2025}. In studying the effects of SMBH mergers, we make the simplifying assumption that the comoving merger rate is constant after $z=10$, and zero before that. For stellar-mass black hole mergers, we assume the comoving rate is zero before $z=15$, and constant after that. A more accurate treatment would include redshift dependence, though this complication is ultimately unnecessary, given that we find black hole mergers are very far from being viable for resolving the tension. 

We assume that each merger converts a fraction $\epsilon = 0.08$ of the total mass ($\rm{M}_{\rm{BINARY}} = 2 \times 10^7 \rm{M}_\odot$ for SMBH binaries, or $\rm{M}_{\rm{BINARY}} = 60 \rm{M}_\odot$ for stellar-mass binaries) to gravitational radiation. In reality, this fraction can vary somewhat, though $\epsilon = 0.08$ is a reasonable approximation, given simulations that show a maximum of $\epsilon \approx  0.11$ for equal-mass binaries with aligned, maximal spins \citep{Healy_2014} and LIGO observations that show $\epsilon \approx 0.05$ \citep{MassLossGW}. Other studies further support this approximation \citep{Barausse_2012}. This yields the following equations (derived from the perfect fluid equation), where $R$ is a comoving merger rate.

\begin{equation}
    \frac{d\rho_m}{dz} = \frac{1}{(1+z)H}(3H\rho_m+R(1+z)^3 \epsilon M_{\rm{BINARY}})
\end{equation}

\begin{equation}
    \frac{d\rho_r}{dz} = \frac{1}{(1+z)H}(4H\rho_m - R(1+z)^3 \epsilon M_{\rm{BINARY}})
\end{equation}

\begin{equation}
    H^2 = \frac{8 \pi G}{3}(\rho_m + \rho_r + \rho_{\Lambda})
\end{equation}

One might initially expect a term of the form $\Omega_{\rm{SMBH}} / \Omega_m$, though this is not needed. The important thing is the rate that matter is converted to radiation, which is set by the merger rate, not the fraction of matter held in SMBHs. One could argue that the magnitude of $R$ is indirectly related to $\Omega_{\rm{SMBH}}$ by the requirement that there exist sufficient SMBHs to undergo that many mergers, though $\Omega_{\rm{SMBH}}$ is not the quantity that is directly relevant. To see why a term $\propto \Omega_{\rm{SMBH}} / \Omega_m$ is not needed, consider that there may be many SMBHs that never undergo mergers, and thus are not involved in the conversion of matter to radiation via mergers. One could modify the analysis to evolve SMBHs separately from the rest of the matter, thereby splitting the matter equation into two coupled differential equations, though this would unnecessarily complicate things, as one would have to know SMBH formation and accretion rates. 

We integrate this system from $z_{REC}=1090$ to $z=0$. In our pipeline, we have two free parameters: 

\begin{enumerate}
    \item $h_i$, which sets the initial densities at $z=1090$
    \item Either a comoving merger rate, or SMBH formation rate, which sets the rate that matter is converted to gravitational radiation.
\end{enumerate}

\noindent No other quantities are varied. Initial densities follow from equations~\eqref{eq:OmegaR} -- \eqref{eq:RhoLambda}, where our method implies $\rho_{m_{initial}} / \rho_{r_{initial}} \approx 3.15$. 

We emphasize that $h_i$ is just a parameter used to set the initial conditions at $z=1090$. Due to the conversion of matter to radiation, $h_i$ does not equal $h$ found from evolving the system to $z=0$. This then implies that varying initial conditions via $h_i$ is not directly equivalent to varying $\rm{H}_0 = \rm{H}(z=0)$ and $\Omega_m (z=0)$. The quantities $\Omega_{m_{initial}}$, $\Omega_{r_{initial}}$, and $\Omega_{{\Lambda}_{initial}}$ produced by the equations below do not exactly match $\Omega_m (z=0)$, $\Omega_r (z=0)$, and $\Omega_{\Lambda} (z=0)$, as some of the matter is converted to gravitational radiation. The quantities $\Omega_m (z=0)$, $\Omega_r (z=0)$, $\Omega_{\Lambda} (z=0)$, and $\rm{H}_0$ are instead indirectly determined by the fact that the system of coupled differential equations has a unique solution for a given set of initial conditions. This is very similar to what was done in past studies that looked at decaying dark matter \citep{DM_Decay, DM_Decay_Constrained}. Also note that $\rm{H_{0_{SI}}}$ is $100 \rm{\frac{km}{sec Mpc}} h_i$ converted to units of $\rm{sec^{-1}}$. The radiation component is governed by the equation below \cite{Dodelson}.
\begin{equation}
    \Omega_{r_{initial}} = 4.15 \times 10^{-5} /h_i^2
    \label{eq:OmegaR}
\end{equation}

\noindent The matter component is governed by the equation below \cite{Planck_H0}. The following equations then provide the densities. 
\begin{equation}
    \Omega_{m_{initial}} = 0.14241/h_i^2
    \label{eq:OmegaM}
\end{equation}

\begin{equation}
    \rho_{CRIT_0} = \rm{3H^2_{0_{SI}}}/8 \pi G
    \label{eq:RhoCrit0}
\end{equation}

\begin{equation}
    \Omega_{\Lambda_{initial}} = 1.0 - \Omega_{m_{initial}} - \Omega_{r_{initial}}
    \label{eq:OmegaLambda}
\end{equation}

\begin{equation}
    \rho_{m_{initial}} = \Omega_{m_{initial}} \rho_{CRIT_0} (1 + z_{REC})^3
    \label{eq:RhoM}
\end{equation}

\begin{equation}
    \rho_{r_{initial}} = \Omega_{r_{initial}} \rho_{CRIT_0} (1 + z_{REC})^4
    \label{eq:RhoR}
\end{equation}

\begin{equation}
    \rho_{\Lambda_{initial}} = \Omega_{\Lambda_{initial}} \rho_{CRIT_0} 
    \label{eq:RhoLambda}
\end{equation}

We then vary $h_i$ and the comoving merger rate to fit to data for late-time $\rm{H(z)}$ \citep{Local_H0, HVal1, HVal2, HVal3, HVal4}. These were the same values for $\rm{H(z)}$ used in \cite{DM_Decay}. We use the L-BFGS-B algorithm packaged within SciPy to minimize $\chi^2$ \citep{Minimizer}. To produce the initial guesses for the minimizer, we first conduct a brute-force grid search over the parameter space to find the parameter combination with the smallest $\chi^2$. This approach is viable here because we only have two free parameters. For the grid search, we try 20 linearly spaced values for $h_i$ between 0.5 and 1. For SMBH mergers, we try 100 linearly spaced values between 0 and $1 \times 10^4$ $\rm{Gpc}^{-3} yr^{-1}$ for $R$. For stellar-mass mergers, we try 100 linearly spaced values between 0 and $3.33 \times 10^9$ $\rm{Gpc}^{-3} yr^{-1}$ for $R$. (This is the same range as for SMBH mergers, scaled by $\frac{10^7}{30}$.) For SMBH formation from PBHs, we are interested in the constant in the numerator of the formation rate function (see below). We try 100 linearly spaced values between 0 and 100 for C.  

\begin{figure}[h!]
    \centerline{
    \includegraphics[width=8.5 cm]{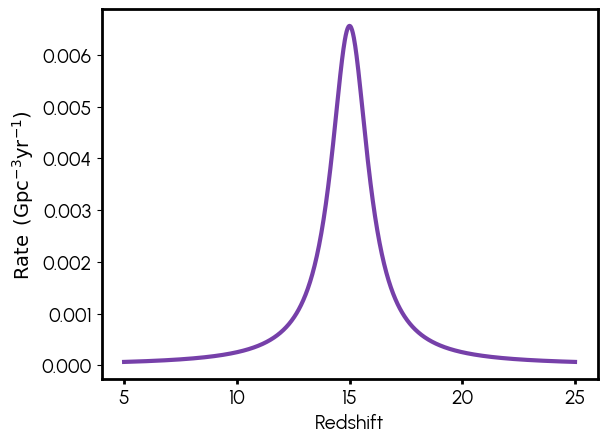}
    }
    \caption{Here we plot the comoving SMBH formation rate in units of $\rm{Gpc}^{-3} yr^{-1}$, that results in a number density of $10^6$ SMBHs per $\rm{Gpc}^{3}$ at $z=0$.} 
    \label{fig:SMBH_Formation_Rate}
\end{figure}

When considering SMBH formation from PBHs, we follow a similar process, though instead of a comoving merger rate, we consider a comoving SMBH formation rate of the following form, plotted in Figure \ref{fig:SMBH_Formation_Rate}.

\begin{equation}
    \rm{Rate} = \frac{C}{1+(z-15)^2}
\end{equation}

This is intended as a simple toy model for a burst of SMBH formation. While the exact functional form is somewhat arbitrary (one could perhaps consider other unimodal functions, such as a Gaussian), the effect on the Hubble tension is driven largely by the fraction of matter in the Universe that is converted to radiation, which is determined by the integral of this rate, instead of its precise shape. We consider this simple model to be sufficient for a rough estimate of the effect. 

As a toy model, we consider a cloud of $10^{18}$ kg PBHs, where we assume that the black holes merge in pairs so that the total number of black holes is cut in half in each round of mergers. If the final black hole mass is $10^7 \rm{M}_\odot$, then this implies that 31 times the final black hole mass ($10^7 \rm{M}_\odot$) is converted to gravitational radiation, when each merger converts five percent of the mass of the binary to radiation. If each merger converts eight percent of the mass of the binary to gravitational radiation, 332 times the mass of the final black hole ($10^7 \rm{M}_\odot$) is converted to radiation. Since we are looking for an order of magnitude estimate, we say that 100 times the mass of the SMBH is converted to radiation, for every SMBH that forms. 

Such a scenario would likely take a prohibitively long time to form a SMBH, though we are not concerned with this, as it is not meant to be a realistic formation scenario. This is meant to illustrate that even an extreme example is not a viable solution to the Hubble tension, as too little matter is converted to radiation. This naturally implies that more realistic scenarios for SMBH formation from mergers are also not viable solutions, as they would convert far less matter to radiation. 

\FloatBarrier

\FloatBarrier

\subsection{\label{sec:ISWEstimation}Estimating the ISW Effect}

When black holes merge, they convert some of the mass of the binary to radiation. Similarly to decaying dark matter, if enough matter is converted to radiation, dark energy will dominate sooner, and the CMB will be boosted at low multipole moment via the ISW effect. This is known to constrain models of decaying dark matter \citep{DM_Decay_Constrained}, so we seek to determine whether this effect can be used to constrain scenarios for SMBH formation from mergers of smaller black holes. When we constrain the comoving SMBH formation rate so that SMBHs are not overproduced, we find that a very small fraction of the matter in the Universe is converted to radiation, even when SMBHs are produced from PBHs. The conversion of matter to radiation then has a negligible effect on $\rm{H(z)}$ when SMBHs are not overproduced, so any effect on $\rm{H(z)}$ is due to the adjustment to the initial conditions by setting $h_i$ when fitting to experimentally measured $\rm{H(z)}$. We can then approximate the effect on the CMB without modifying the CLASS package \cite{CLASS_Package}, where we simply adjust $h$, setting $h = h_i$, since this is ignoring the minuscule effects from converting matter to radiation. The effect on the CMB is so small that an approximate result is sufficient. 

\section{\label{sec:resuls} Results and Discussion}

\subsection{\label{sec:SMBHMergers}SMBH Mergers}

\begin{figure}[h!]
    \centerline{
    \includegraphics[width=8.5 cm]{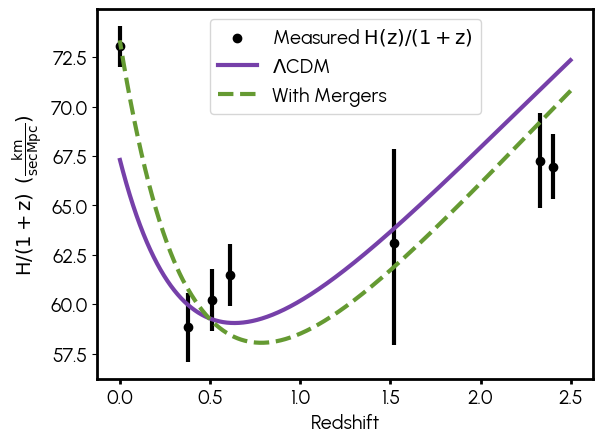}
    }
    \caption{Here we show the evolution of the Hubble parameter with redshift, with and without SMBH mergers, with the best-fit comoving merger rate of around 2600 $\rm{Gpc}^{-3} \rm{yr}^{-1}$. This is unreasonably large, making this mechanism for alleviating the tension non-viable.} 
    \label{fig:SMBH_Merger_Effects_H}
\end{figure}

\begin{figure}[h!]
    \centerline{
    \includegraphics[width=8.5 cm]{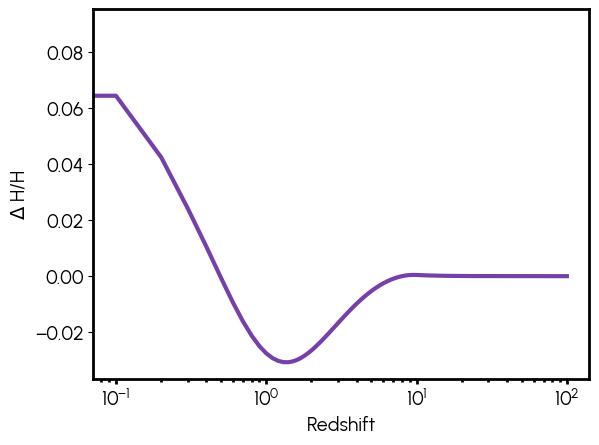}
    }
    \caption{Here we show the fractional change of the Hubble parameter with SMBH mergers as compared to the case without mergers. This is using the best-fit comoving merger rate of around 2600 $\rm{Gpc}^{-3} \rm{yr}^{-1}$. One can readily see that the case with mergers has a larger Hubble parameter at late times.} 
    \label{fig:SMBH_Merger_Effects_Delta_H}
\end{figure}

\begin{figure}[h!]
    \centerline{
    \includegraphics[width=8.5 cm]{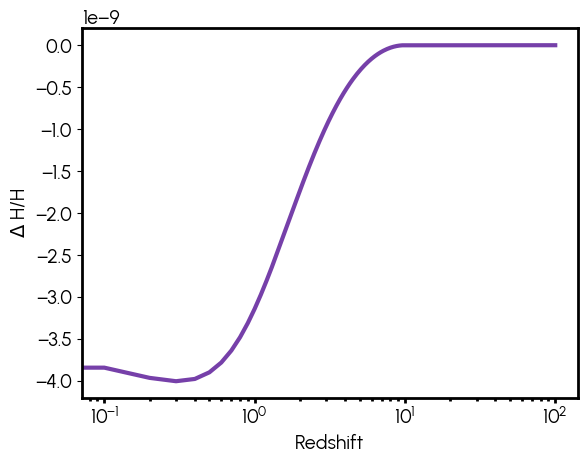}
    }
    \caption{Here we show the fractional change of the Hubble parameter (with a more realistic SMBH merger rate, here approximated as a constant $10^{-4}$ $\rm{Gpc}^{-3} \rm{yr}^{-1}$) as compared to the case without mergers. To isolate the effects of the mergers from the effects of varying $h_i$, for this plot, we fix $h_{i} = 0.674$, yielding $\Omega_{m_{initial}} \approx \Omega_m(z=0) \approx 0.313$. We have plotted the fractional change in the Hubble parameter rather than $\rm{H(z)}$ because the effect is so small that it would be invisible on an $\rm{H(z)}$ plot.} 
    \label{fig:Realistic_SMBH_Merger_Rate}
\end{figure}

We find that the data favors a comoving merger rate of approximately 2600 $\rm{Gpc}^{-3} \rm{yr}^{-1}$, with initial conditions following from $h_i \approx 0.786$, which corresponds to $\Omega_{m_{initial}} \approx 0.23$. This yields $\rm{H_0} \approx$ 73.4 km  $\rm{sec}^{-1} \rm{Mpc}^{-1}$ and $\Omega_m (z=0) \approx 0$. The fit is driven towards a merger rate so extreme it would convert nearly all matter in the Universe into gravitational radiation, highlighting how unrealistic such a merger rate is. While SMBH merger rates are not known for certain, this is at least a few orders of magnitude larger than the expected rate \citep{SMBH_Merger_Rate}, which suggests that SMBH mergers are not the solution to the Hubble tension. One should also note that pulsar timing array results suggest a merger rate of around $1.2 \times 10^{-5}$ to $4.5 \times 10^{-4}$ $\rm{Gpc}^{-3} \rm{yr}^{-1}$ \citep{Middleton_2021}. We show the effect on $\rm{H(z)}$ in Figure \ref{fig:SMBH_Merger_Effects_H} and Figure \ref{fig:SMBH_Merger_Effects_Delta_H}. For comparison, we also illustrate the effect of a more realistic merger rate in Figure \ref{fig:Realistic_SMBH_Merger_Rate}. 

\FloatBarrier

\FloatBarrier

\subsection{\label{sec:SMBHFormationPBHs}SMBH Formation from PBHs}

\begin{figure}[h!]
    \centerline{
    \includegraphics[width=8.5 cm]{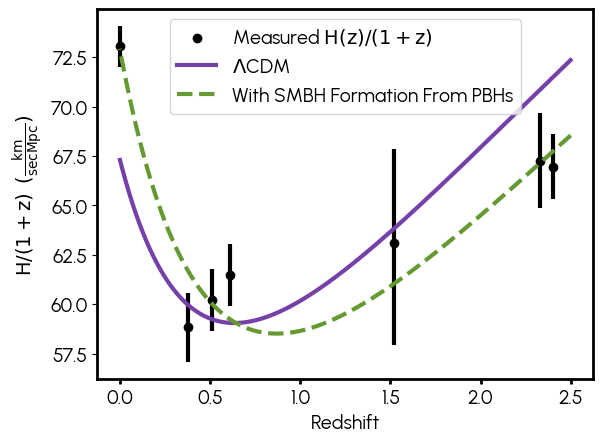}
    }
    \caption{Here we show the evolution of the Hubble parameter with redshift, with and without SMBH formation from PBHs, with the best-fit comoving formation rate. This best-fit rate can alleviate the Hubble tension, but overproduces SMBHs by three to four orders of magnitude.} 
    \label{fig:SMBH_Formation_H}
\end{figure}

\begin{figure}[h!]
    \centerline{
    \includegraphics[width=8.5 cm]{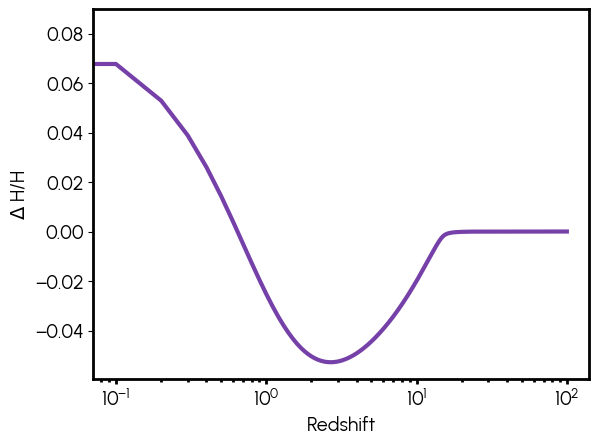}
    }
    \caption{Here we show the fractional change of the Hubble parameter with redshift, with and without SMBH formation from PBHs, with the best-fit comoving formation rate. This best-fit rate can alleviate the Hubble tension, but overproduces SMBHs by three to four orders of magnitude.} 
    \label{fig:SMBH_Formation_Delta_H}
\end{figure}

SMBH Formation from PBHs is an extreme scenario, though it is useful for illustrating that even such a case is unlikely to alleviate the Hubble tension, or be constrained via the ISW effect. We find that in order to alleviate the Hubble tension, one would need to overproduce SMBHs by a factor of around $10^3$ to $10^4$, as our Universe is believed to have a SMBH number density of around $10^6$ to $10^7$ $\rm{Gpc^{-3}}$ at $z=0$ \citep{SMBHNumberDensity}, so far too little matter would be converted to radiation, if SMBHs are not overproduced. We find a best-fit value of around 74.1 $\rm{Gpc^{-3} yr^{-1}}$ for the constant in the numerator of the comoving formation rate function, with initial conditions following from $h_i \approx 0.750$, corresponding to $\Omega_{m_{initial}} \approx 0.25$. This yields $\rm{H_0} \approx$ 72.9 km  $\rm{sec}^{-1} \rm{Mpc}^{-1}$ and $\Omega_m (z=0) \approx 0.19$. This is less severe than the case of SMBH mergers, in that the fit is not driven to destroy all of the matter in the Universe, though the fraction of matter converted to radiation is still incredibly unrealistic, as this is roughly equivalent to converting all of the baryons in the Universe to gravitational radiation. We show the impacts on \rm{H(z)} in Figure \ref{fig:SMBH_Formation_H} and Figure \ref{fig:SMBH_Formation_Delta_H}. 

We also consider the effect on the CMB via the ISW effect where the SMBH formation rate is constrained so that SMBHs are not overproduced. We find that the boost to the CMB power spectrum at low multipole moment is around five percent, which suggests that the effect is well within the error bars on the CMB. One should note that because a tiny fraction of the matter in the Universe is converted to radiation when SMBHs are not overproduced, the small boost here is entirely due to the fit preferring a larger value for $h_i$ ($h_i \approx 0.7185$), which works to shift the initial conditions at $z=1090$. Given that very little matter can be converted to radiation in this constrained scenario, the fit can only match empirically measured H(z) by increasing $\Omega_{\Lambda_{initial}}$ through raising $h_i$. This value for $h_i$ corresponds to $\Omega_{m_{initial}} \approx \Omega_{m} (z=0) \approx 0.28$ and $\Omega_{\Lambda_{initial}} \approx \Omega_{\Lambda}(z=0) \approx 0.72$. We see that the ISW effect likely cannot be used to constrain realistic SMBH formation channels, as realistic channels would convert far less matter to radiation. 

\FloatBarrier

\FloatBarrier

\subsection{\label{sec:StellarBHs}Mergers of Stellar-Mass Black Holes}

\FloatBarrier

\begin{figure}[h!]
    \centerline{
    \includegraphics[width=8.5 cm]{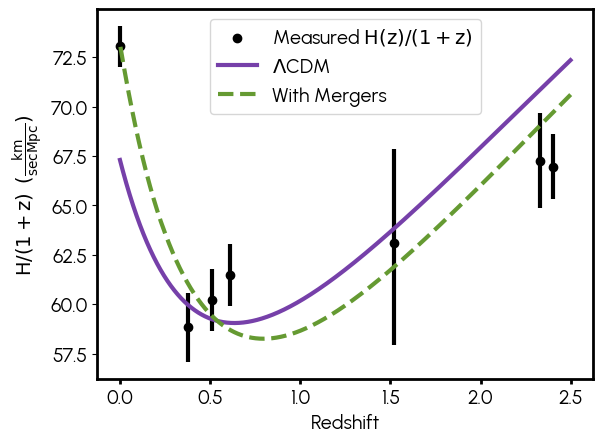}
    }
    \caption{Here we show the evolution of the Hubble parameter with redshift, with and without mergers of stellar-mass black holes (30 $\rm{M}_\odot$), with the best-fit comoving merger rate. This best-fit rate can alleviate the Hubble tension, but requires an unphysically large merger rate.} 
    \label{fig:Stellar_BH_H}
\end{figure}

\begin{figure}[h!]
    \centerline{
    \includegraphics[width=8.5 cm]{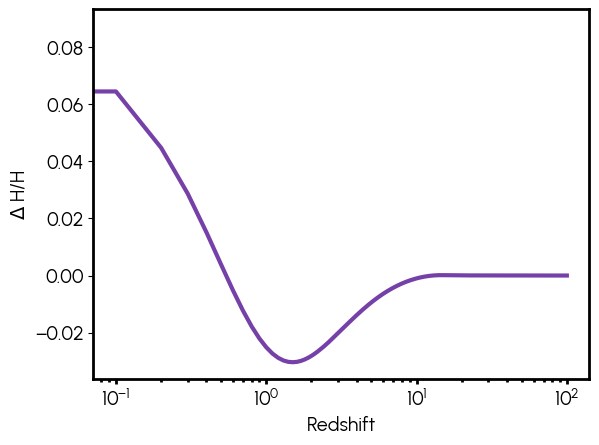}
    }
    \caption{Here we show the fractional change of the Hubble parameter with redshift, with and without mergers of stellar-mass black holes (30 $\rm{M}_\odot$), with the best-fit comoving merger rate. This best-fit rate can alleviate the Hubble tension, but requires an unphysically large merger rate.} 
    \label{fig:Stellar_Delta_H}
\end{figure}

\begin{figure}[H]
    \centerline{
    \includegraphics[width=8.5 cm]{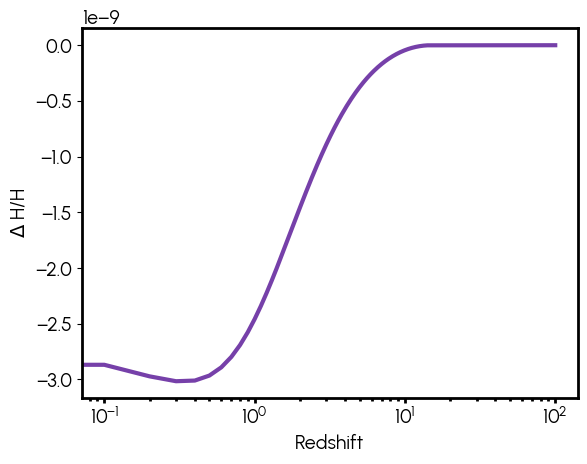}
    }
    \caption{Here we show the fractional change of the Hubble parameter as compared to the case without mergers. This uses a more realistic stellar-mass black hole merger rate \cite{LIGO_Merger_Rate}, here approximated as a constant 24 $\rm{Gpc}^{-3} \rm{yr}^{-1}$. To isolate the effects of the mergers from the effects of varying $h_i$, for this plot, we fix $h_{i} = 0.674$, yielding $\Omega_{m_{initial}} \approx \Omega_m(z=0) \approx 0.313$. We have plotted the fractional change in the Hubble parameter rather than $\rm{H(z)}$ because the effect is so small that it would be invisible on an $\rm{H(z)}$ plot.} 
    \label{fig:Realistic_Solar_Merger_Rate}
\end{figure}

We also consider mergers of stellar-mass black holes, with masses of 30 $\rm{M}_\odot$. This is roughly in line with the masses commonly seen in LIGO events. Much like the case of SMBH mergers, we find that the necessary merger rate for alleviating the Hubble tension (approx. $7.6 \times 10^8$ $\rm{Gpc^{-3}} \rm{yr^{-1}}$) is many orders of magnitude larger than the rate inferred from LIGO events (approx. 24 $\rm{Gpc^{-3}} \rm{yr^{-1}}$) \cite{LIGO_Merger_Rate}. Our best-fit initial conditions follow from $h_i \approx 0.782$, which corresponds to $\Omega_{m_{initial}} \approx 0.23$. This yields $\rm{H}_0 \approx 73.2$ km  $\rm{sec}^{-1} \rm{Mpc}^{-1}$ and $\Omega_m (z=0) \approx 0$. As with SMBH mergers, this is incredibly unrealistic, as the fit is driven to such an extreme merger rate that it would convert nearly all the matter in the Universe to gravitational radiation. We show the impacts on \rm{H(z)} in Figure \ref{fig:Stellar_BH_H} and Figure \ref{fig:Stellar_Delta_H}. For comparison, we also show the impact of a more realistic merger rate in Figure \ref{fig:Realistic_Solar_Merger_Rate}. 

\FloatBarrier

\FloatBarrier

\FloatBarrier

\FloatBarrier

\newpage

\section{\label{sec:conclusion}Conclusion}

We find that neither SMBH mergers nor stellar-mass black hole mergers can alleviate the Hubble tension, as the required merger rates are unreasonably large. We find that formation of SMBHs from mergers of smaller black holes also cannot alleviate the tension, as one would need to overproduce SMBHs. We also see that SMBH formation channels via mergers cannot be easily constrained using the ISW effect, as too little matter is converted to radiation. Our work perhaps further suggests that the Hubble tension could be a sign of new physics. 

\begin{acknowledgments}
We thank Bob Scherrer for helpful discussions.
\end{acknowledgments}

\bibliography{apssamp}

\begin{thebibliography}{30}%
\makeatletter
\providecommand \@ifxundefined [1]{%
 \@ifx{#1\undefined}
}%
\providecommand \@ifnum [1]{%
 \ifnum #1\expandafter \@firstoftwo
 \else \expandafter \@secondoftwo
 \fi
}%
\providecommand \@ifx [1]{%
 \ifx #1\expandafter \@firstoftwo
 \else \expandafter \@secondoftwo
 \fi
}%
\providecommand \natexlab [1]{#1}%
\providecommand \enquote  [1]{``#1''}%
\providecommand \bibnamefont  [1]{#1}%
\providecommand \bibfnamefont [1]{#1}%
\providecommand \citenamefont [1]{#1}%
\providecommand \href@noop [0]{\@secondoftwo}%
\providecommand \href [0]{\begingroup \@sanitize@url \@href}%
\providecommand \@href[1]{\@@startlink{#1}\@@href}%
\providecommand \@@href[1]{\endgroup#1\@@endlink}%
\providecommand \@sanitize@url [0]{\catcode `\\12\catcode `\$12\catcode `\&12\catcode `\#12\catcode `\^12\catcode `\_12\catcode `\%12\relax}%
\providecommand \@@startlink[1]{}%
\providecommand \@@endlink[0]{}%
\providecommand \url  [0]{\begingroup\@sanitize@url \@url }%
\providecommand \@url [1]{\endgroup\@href {#1}{\urlprefix }}%
\providecommand \urlprefix  [0]{URL }%
\providecommand \Eprint [0]{\href }%
\providecommand \doibase [0]{https://doi.org/}%
\providecommand \selectlanguage [0]{\@gobble}%
\providecommand \bibinfo  [0]{\@secondoftwo}%
\providecommand \bibfield  [0]{\@secondoftwo}%
\providecommand \translation [1]{[#1]}%
\providecommand \BibitemOpen [0]{}%
\providecommand \bibitemStop [0]{}%
\providecommand \bibitemNoStop [0]{.\EOS\space}%
\providecommand \EOS [0]{\spacefactor3000\relax}%
\providecommand \BibitemShut  [1]{\csname bibitem#1\endcsname}%
\let\auto@bib@innerbib\@empty
\bibitem [{\citenamefont {Hubble}(1929)}]{Hubble}%
  \BibitemOpen
  \bibfield  {author} {\bibinfo {author} {\bibfnamefont {E.}~\bibnamefont {Hubble}},\ }\bibfield  {title} {\bibinfo {title} {A relation between distance and radial velocity among extra-galactic nebulae},\ }\href {https://doi.org/10.1073/pnas.15.3.168} {\bibfield  {journal} {\bibinfo  {journal} {Proceedings of the National Academy of Sciences}\ }\textbf {\bibinfo {volume} {15}},\ \bibinfo {pages} {168} (\bibinfo {year} {1929})},\ \Eprint {https://arxiv.org/abs/https://www.pnas.org/doi/pdf/10.1073/pnas.15.3.168} {https://www.pnas.org/doi/pdf/10.1073/pnas.15.3.168} \BibitemShut {NoStop}%
\bibitem [{\citenamefont {Carroll}\ and\ \citenamefont {Ostlie}(2017)}]{ModernAstrophysicsBook}%
  \BibitemOpen
  \bibfield  {author} {\bibinfo {author} {\bibfnamefont {B.}~\bibnamefont {Carroll}}\ and\ \bibinfo {author} {\bibfnamefont {D.}~\bibnamefont {Ostlie}},\ }\bibinfo {title} {An introduction to modern astrophysics, 2nd edition}\ (\bibinfo  {publisher} {Cambridge University Press},\ \bibinfo {year} {2017})\BibitemShut {NoStop}%
\bibitem [{\citenamefont {Maguire}(2017)}]{Maguire2017}%
  \BibitemOpen
  \bibfield  {author} {\bibinfo {author} {\bibfnamefont {K.}~\bibnamefont {Maguire}},\ }\bibinfo {title} {Type ia supernovae},\ in\ \href {https://doi.org/10.1007/978-3-319-21846-5_36} {\emph {\bibinfo {booktitle} {Handbook of Supernovae}}},\ \bibinfo {editor} {edited by\ \bibinfo {editor} {\bibfnamefont {A.~W.}\ \bibnamefont {Alsabti}}\ and\ \bibinfo {editor} {\bibfnamefont {P.}~\bibnamefont {Murdin}}}\ (\bibinfo  {publisher} {Springer International Publishing},\ \bibinfo {address} {Cham},\ \bibinfo {year} {2017})\ pp.\ \bibinfo {pages} {293--316}\BibitemShut {NoStop}%
\bibitem [{\citenamefont {Riess}\ \emph {et~al.}(2022)\citenamefont {Riess} \emph {et~al.}}]{Local_H0}%
  \BibitemOpen
  \bibfield  {author} {\bibinfo {author} {\bibfnamefont {A.~G.}\ \bibnamefont {Riess}} \emph {et~al.},\ }\bibfield  {title} {\bibinfo {title} {A comprehensive measurement of the local value of the hubble constant with 1 km $\rm{s^{-1}}$ $\rm{Mpc^{-1}}$ uncertainty from the hubble space telescope and the sh0es team},\ }\href {https://doi.org/10.3847/2041-8213/ac5c5b} {\bibfield  {journal} {\bibinfo  {journal} {The Astrophysical Journal Letters}\ }\textbf {\bibinfo {volume} {934}},\ \bibinfo {pages} {L7} (\bibinfo {year} {2022})}\BibitemShut {NoStop}%
\bibitem [{\citenamefont {Kamionkowski}\ and\ \citenamefont {Riess}(2023)}]{Early_Dark_Energy}%
  \BibitemOpen
  \bibfield  {author} {\bibinfo {author} {\bibfnamefont {M.}~\bibnamefont {Kamionkowski}}\ and\ \bibinfo {author} {\bibfnamefont {A.~G.}\ \bibnamefont {Riess}},\ }\bibfield  {title} {\bibinfo {title} {The hubble tension and early dark energy},\ }\href {https://doi.org/https://doi.org/10.1146/annurev-nucl-111422-024107} {\bibfield  {journal} {\bibinfo  {journal} {Annual Review of Nuclear and Particle Science}\ }\textbf {\bibinfo {volume} {73}},\ \bibinfo {pages} {153} (\bibinfo {year} {2023})}\BibitemShut {NoStop}%
\bibitem [{\citenamefont {Durrer}(2015)}]{CMB_Review}%
  \BibitemOpen
  \bibfield  {author} {\bibinfo {author} {\bibfnamefont {R.}~\bibnamefont {Durrer}},\ }\bibfield  {title} {\bibinfo {title} {The cosmic microwave background: the history of its experimental investigation and its significance for cosmology},\ }\bibfield  {journal} {\bibinfo  {journal} {Classical and Quantum Gravity}\ }\textbf {\bibinfo {volume} {32}},\ \href {https://doi.org/10.1088/0264-9381/32/12/124007} {10.1088/0264-9381/32/12/124007} (\bibinfo {year} {2015})\BibitemShut {NoStop}%
\bibitem [{\citenamefont {Aghanim}\ \emph {et~al.}(2020)\citenamefont {Aghanim} \emph {et~al.}}]{Planck_H0}%
  \BibitemOpen
  \bibfield  {author} {\bibinfo {author} {\bibfnamefont {N.}~\bibnamefont {Aghanim}} \emph {et~al.},\ }\bibfield  {title} {\bibinfo {title} {Planck 2018 results - vi. cosmological parameters},\ }\href {https://doi.org/10.1051/0004-6361/201833910} {\bibfield  {journal} {\bibinfo  {journal} {Astronomy and Astrophysics}\ }\textbf {\bibinfo {volume} {641}},\ \bibinfo {pages} {A6} (\bibinfo {year} {2020})}\BibitemShut {NoStop}%
\bibitem [{\citenamefont {Vattis}\ \emph {et~al.}(2019)\citenamefont {Vattis}, \citenamefont {Koushiappas},\ and\ \citenamefont {Loeb}}]{DM_Decay}%
  \BibitemOpen
  \bibfield  {author} {\bibinfo {author} {\bibfnamefont {K.}~\bibnamefont {Vattis}}, \bibinfo {author} {\bibfnamefont {S.~M.}\ \bibnamefont {Koushiappas}},\ and\ \bibinfo {author} {\bibfnamefont {A.}~\bibnamefont {Loeb}},\ }\bibfield  {title} {\bibinfo {title} {Dark matter decaying in the late universe can relieve the $\rm{{H}_{0}}$ tension},\ }\bibfield  {journal} {\bibinfo  {journal} {Phys. Rev. D}\ }\textbf {\bibinfo {volume} {99}},\ \href {https://doi.org/10.1103/PhysRevD.99.121302} {10.1103/PhysRevD.99.121302} (\bibinfo {year} {2019})\BibitemShut {NoStop}%
\bibitem [{\citenamefont {Pandey}\ \emph {et~al.}(2020)\citenamefont {Pandey}, \citenamefont {Karwal},\ and\ \citenamefont {Das}}]{Pandey_2020}%
  \BibitemOpen
  \bibfield  {author} {\bibinfo {author} {\bibfnamefont {K.~L.}\ \bibnamefont {Pandey}}, \bibinfo {author} {\bibfnamefont {T.}~\bibnamefont {Karwal}},\ and\ \bibinfo {author} {\bibfnamefont {S.}~\bibnamefont {Das}},\ }\bibfield  {title} {\bibinfo {title} {Alleviating the $\rm{H_0}$ and $\sigma_8$ anomalies with a decaying dark matter model},\ }\href {https://doi.org/10.1088/1475-7516/2020/07/026} {\bibfield  {journal} {\bibinfo  {journal} {Journal of Cosmology and Astroparticle Physics}\ }\textbf {\bibinfo {volume} {2020}}\bibinfo  {number} { (07)}}\BibitemShut {NoStop}%
\bibitem [{\citenamefont {Rubin}\ and\ \citenamefont {Ford}(1970)}]{Rubin}%
  \BibitemOpen
\bibfield  {number} {  }\bibfield  {author} {\bibinfo {author} {\bibfnamefont {V.~C.}\ \bibnamefont {Rubin}}\ and\ \bibinfo {author} {\bibfnamefont {J.}~\bibnamefont {Ford}, \bibfnamefont {W.~Kent}},\ }\bibfield  {title} {\bibinfo {title} {Rotation of the andromeda nebula from a spectroscopic survey of emission regions},\ }\href@noop {} {\bibfield  {journal} {\bibinfo  {journal} {\apj}\ }\textbf {\bibinfo {volume} {159}},\ \bibinfo {pages} {379} (\bibinfo {year} {1970})}\BibitemShut {NoStop}%
\bibitem [{\citenamefont {Arbey}\ and\ \citenamefont {Mahmoudi}(2021)}]{DMReview}%
  \BibitemOpen
  \bibfield  {author} {\bibinfo {author} {\bibfnamefont {A.}~\bibnamefont {Arbey}}\ and\ \bibinfo {author} {\bibfnamefont {F.}~\bibnamefont {Mahmoudi}},\ }\bibfield  {title} {\bibinfo {title} {Dark matter and the early universe: A review},\ }\href@noop {} {\bibfield  {journal} {\bibinfo  {journal} {Progress in Particle and Nuclear Physics}\ }\textbf {\bibinfo {volume} {119}} (\bibinfo {year} {2021})}\BibitemShut {NoStop}%
\bibitem [{\citenamefont {Abbott}\ \emph {et~al.}(2016)\citenamefont {Abbott} \emph {et~al.}}]{MassLossGW}%
  \BibitemOpen
  \bibfield  {author} {\bibinfo {author} {\bibfnamefont {B.~P.}\ \bibnamefont {Abbott}} \emph {et~al.} (\bibinfo {collaboration} {LIGO Scientific Collaboration and Virgo Collaboration}),\ }\bibfield  {title} {\bibinfo {title} {Observation of gravitational waves from a binary black hole merger},\ }\href {https://doi.org/10.1103/PhysRevLett.116.061102} {\bibfield  {journal} {\bibinfo  {journal} {Phys. Rev. Lett.}\ }\textbf {\bibinfo {volume} {116}},\ \bibinfo {pages} {061102} (\bibinfo {year} {2016})}\BibitemShut {NoStop}%
\bibitem [{\citenamefont {Eroshenko}(2021)}]{PBH}%
  \BibitemOpen
  \bibfield  {author} {\bibinfo {author} {\bibfnamefont {Y.}~\bibnamefont {Eroshenko}},\ }\bibfield  {title} {\bibinfo {title} {Mergers of primordial black holes in extreme clusters and the $\rm{H}_0$ tension},\ }\bibfield  {journal} {\bibinfo  {journal} {Physics of the Dark Universe}\ }\textbf {\bibinfo {volume} {32}},\ \href {https://doi.org/https://doi.org/10.1016/j.dark.2021.100833} {https://doi.org/10.1016/j.dark.2021.100833} (\bibinfo {year} {2021})\BibitemShut {NoStop}%
\bibitem [{\citenamefont {Volonteri}(2010)}]{SMBH_Formation}%
  \BibitemOpen
  \bibfield  {author} {\bibinfo {author} {\bibfnamefont {M.}~\bibnamefont {Volonteri}},\ }\bibfield  {title} {\bibinfo {title} {Formation of supermassive black holes},\ }\bibfield  {journal} {\bibinfo  {journal} {The Astronomy and Astrophysics Review}\ }\textbf {\bibinfo {volume} {18}},\ \href {https://doi.org/10.1007/s00159-010-0029-x} {10.1007/s00159-010-0029-x} (\bibinfo {year} {2010})\BibitemShut {NoStop}%
\bibitem [{\citenamefont {Camarena}\ and\ \citenamefont {Marra}(2016)}]{StructureFormation}%
  \BibitemOpen
  \bibfield  {author} {\bibinfo {author} {\bibfnamefont {D.}~\bibnamefont {Camarena}}\ and\ \bibinfo {author} {\bibfnamefont {V.}~\bibnamefont {Marra}},\ }\bibfield  {title} {\bibinfo {title} {Cosmological constraints on the radiation released during structure formation},\ }\bibfield  {journal} {\bibinfo  {journal} {The European Physical Journal C}\ }\textbf {\bibinfo {volume} {76}},\ \href {https://doi.org/10.1140/epjc/s10052-016-4517-7} {10.1140/epjc/s10052-016-4517-7} (\bibinfo {year} {2016})\BibitemShut {NoStop}%
\bibitem [{\citenamefont {Clark}\ \emph {et~al.}(2021)\citenamefont {Clark}, \citenamefont {Vattis},\ and\ \citenamefont {Koushiappas}}]{DM_Decay_Constrained}%
  \BibitemOpen
  \bibfield  {author} {\bibinfo {author} {\bibfnamefont {S.~J.}\ \bibnamefont {Clark}}, \bibinfo {author} {\bibfnamefont {K.}~\bibnamefont {Vattis}},\ and\ \bibinfo {author} {\bibfnamefont {S.~M.}\ \bibnamefont {Koushiappas}},\ }\bibfield  {title} {\bibinfo {title} {Cosmological constraints on late-universe decaying dark matter as a solution to the $\rm{{H}_{0}}$ tension},\ }\bibfield  {journal} {\bibinfo  {journal} {Phys. Rev. D}\ }\textbf {\bibinfo {volume} {103}},\ \href {https://doi.org/10.1103/PhysRevD.103.043014} {10.1103/PhysRevD.103.043014} (\bibinfo {year} {2021})\BibitemShut {NoStop}%
\bibitem [{\citenamefont {Hoelscher}(2025)}]{Hoelscher2025}%
  \BibitemOpen
  \bibfield  {author} {\bibinfo {author} {\bibfnamefont {Z.}~\bibnamefont {Hoelscher}},\ }\href {https://doi.org/10.5281/zenodo.17527588} {\bibinfo {title} {{Black\_Hole\_Mergers\_Cosmology}}} (\bibinfo {year} {2025})\BibitemShut {NoStop}%
\bibitem [{\citenamefont {Healy}\ \emph {et~al.}(2014)\citenamefont {Healy}, \citenamefont {Lousto},\ and\ \citenamefont {Zlochower}}]{Healy_2014}%
  \BibitemOpen
  \bibfield  {author} {\bibinfo {author} {\bibfnamefont {J.}~\bibnamefont {Healy}}, \bibinfo {author} {\bibfnamefont {C.~O.}\ \bibnamefont {Lousto}},\ and\ \bibinfo {author} {\bibfnamefont {Y.}~\bibnamefont {Zlochower}},\ }\bibfield  {title} {\bibinfo {title} {Remnant mass, spin, and recoil from spin aligned black-hole binaries},\ }\href {https://doi.org/10.1103/PhysRevD.90.104004} {\bibfield  {journal} {\bibinfo  {journal} {Phys. Rev. D}\ }\textbf {\bibinfo {volume} {90}},\ \bibinfo {pages} {104004} (\bibinfo {year} {2014})}\BibitemShut {NoStop}%
\bibitem [{\citenamefont {Barausse}\ \emph {et~al.}(2012)\citenamefont {Barausse}, \citenamefont {Morozova},\ and\ \citenamefont {Rezzolla}}]{Barausse_2012}%
  \BibitemOpen
  \bibfield  {author} {\bibinfo {author} {\bibfnamefont {E.}~\bibnamefont {Barausse}}, \bibinfo {author} {\bibfnamefont {V.}~\bibnamefont {Morozova}},\ and\ \bibinfo {author} {\bibfnamefont {L.}~\bibnamefont {Rezzolla}},\ }\bibfield  {title} {\bibinfo {title} {On the mass radiated by coalescing black hole binaries},\ }\href {https://doi.org/10.1088/0004-637X/758/1/63} {\bibfield  {journal} {\bibinfo  {journal} {The Astrophysical Journal}\ }\textbf {\bibinfo {volume} {758}},\ \bibinfo {pages} {63} (\bibinfo {year} {2012})}\BibitemShut {NoStop}%
\bibitem [{\citenamefont {Dodelson}(2003)}]{Dodelson}%
  \BibitemOpen
  \bibfield  {author} {\bibinfo {author} {\bibfnamefont {S.}~\bibnamefont {Dodelson}},\ }\bibinfo {title} {Modern cosmology}\ (\bibinfo  {publisher} {Elsevier},\ \bibinfo {year} {2003})\BibitemShut {NoStop}%
\bibitem [{\citenamefont {Alam}\ \emph {et~al.}(2017)\citenamefont {Alam} \emph {et~al.}}]{HVal1}%
  \BibitemOpen
  \bibfield  {author} {\bibinfo {author} {\bibfnamefont {S.}~\bibnamefont {Alam}} \emph {et~al.},\ }\bibfield  {title} {\bibinfo {title} {The clustering of galaxies in the completed sdss-iii baryon oscillation spectroscopic survey: cosmological analysis of the dr12 galaxy sample},\ }\href {https://doi.org/10.1093/mnras/stx721} {\bibfield  {journal} {\bibinfo  {journal} {Monthly Notices of the Royal Astronomical Society}\ }\textbf {\bibinfo {volume} {470}},\ \bibinfo {pages} {2617} (\bibinfo {year} {2017})},\ \Eprint {https://arxiv.org/abs/https://academic.oup.com/mnras/article-pdf/470/3/2617/18315003/stx721.pdf} {https://academic.oup.com/mnras/article-pdf/470/3/2617/18315003/stx721.pdf} \BibitemShut {NoStop}%
\bibitem [{\citenamefont {Zarrouk}\ \emph {et~al.}(2018)\citenamefont {Zarrouk} \emph {et~al.}}]{HVal2}%
  \BibitemOpen
  \bibfield  {author} {\bibinfo {author} {\bibfnamefont {P.}~\bibnamefont {Zarrouk}} \emph {et~al.},\ }\bibfield  {title} {\bibinfo {title} {The clustering of the sdss-iv extended baryon oscillation spectroscopic survey dr14 quasar sample: measurement of the growth rate of structure from the anisotropic correlation function between redshift 0.8 and 2.2},\ }\href {https://doi.org/10.1093/mnras/sty506} {\bibfield  {journal} {\bibinfo  {journal} {Monthly Notices of the Royal Astronomical Society}\ }\textbf {\bibinfo {volume} {477}},\ \bibinfo {pages} {1639} (\bibinfo {year} {2018})},\ \Eprint {https://arxiv.org/abs/https://academic.oup.com/mnras/article-pdf/477/2/1639/25009821/sty506.pdf} {https://academic.oup.com/mnras/article-pdf/477/2/1639/25009821/sty506.pdf} \BibitemShut {NoStop}%
\bibitem [{\citenamefont {Bautista}\ \emph {et~al.}(2017)\citenamefont {Bautista} \emph {et~al.}}]{HVal3}%
  \BibitemOpen
  \bibfield  {author} {\bibinfo {author} {\bibfnamefont {J.~E.}\ \bibnamefont {Bautista}} \emph {et~al.},\ }\bibfield  {title} {\bibinfo {title} {Measurement of baryon acoustic oscillation correlations at z=2.3 with sdss dr12 ly $\alpha$ -forests},\ }\href {https://doi.org/10.1051/0004-6361/201730533} {\bibfield  {journal} {\bibinfo  {journal} {A \& A}\ }\textbf {\bibinfo {volume} {603}},\ \bibinfo {pages} {A12} (\bibinfo {year} {2017})}\BibitemShut {NoStop}%
\bibitem [{\citenamefont {du~Mas~des Bourboux}\ \emph {et~al.}(2017)\citenamefont {du~Mas~des Bourboux} \emph {et~al.}}]{HVal4}%
  \BibitemOpen
  \bibfield  {author} {\bibinfo {author} {\bibfnamefont {H.}~\bibnamefont {du~Mas~des Bourboux}} \emph {et~al.},\ }\bibfield  {title} {\bibinfo {title} {Baryon acoustic oscillations from the complete sdss-iii ly$\alpha$-quasar cross-correlation function at z = 2.4},\ }\bibfield  {journal} {\bibinfo  {journal} {A \& A}\ }\textbf {\bibinfo {volume} {608}},\ \href {https://doi.org/10.1051/0004-6361/201731731} {10.1051/0004-6361/201731731} (\bibinfo {year} {2017})\BibitemShut {NoStop}%
\bibitem [{\citenamefont {Zhu}\ \emph {et~al.}(1997)\citenamefont {Zhu}, \citenamefont {Byrd}, \citenamefont {Lu},\ and\ \citenamefont {Nocedal}}]{Minimizer}%
  \BibitemOpen
  \bibfield  {author} {\bibinfo {author} {\bibfnamefont {C.}~\bibnamefont {Zhu}}, \bibinfo {author} {\bibfnamefont {R.~H.}\ \bibnamefont {Byrd}}, \bibinfo {author} {\bibfnamefont {P.}~\bibnamefont {Lu}},\ and\ \bibinfo {author} {\bibfnamefont {J.}~\bibnamefont {Nocedal}},\ }\bibfield  {title} {\bibinfo {title} {Algorithm 778: L-bfgs-b: Fortran subroutines for large-scale bound-constrained optimization},\ }\href {https://doi.org/10.1145/279232.279236} {\bibfield  {journal} {\bibinfo  {journal} {ACM Trans. Math. Softw.}\ }\textbf {\bibinfo {volume} {23}},\ \bibinfo {pages} {550–560} (\bibinfo {year} {1997})}\BibitemShut {NoStop}%
\bibitem [{\citenamefont {Blas}\ \emph {et~al.}(2011)\citenamefont {Blas}, \citenamefont {Lesgourgues},\ and\ \citenamefont {Tram}}]{CLASS_Package}%
  \BibitemOpen
  \bibfield  {author} {\bibinfo {author} {\bibfnamefont {D.}~\bibnamefont {Blas}}, \bibinfo {author} {\bibfnamefont {J.}~\bibnamefont {Lesgourgues}},\ and\ \bibinfo {author} {\bibfnamefont {T.}~\bibnamefont {Tram}},\ }\bibfield  {title} {\bibinfo {title} {The cosmic linear anisotropy solving system (class). part ii: Approximation schemes},\ }\href {https://doi.org/10.1088/1475-7516/2011/07/034} {\bibfield  {journal} {\bibinfo  {journal} {Journal of Cosmology and Astroparticle Physics}\ }\textbf {\bibinfo {volume} {2011}}\bibinfo  {number} { (07)},\ \bibinfo {pages} {034}}\BibitemShut {NoStop}%
\bibitem [{\citenamefont {Erickcek}\ \emph {et~al.}(2006)\citenamefont {Erickcek}, \citenamefont {Kamionkowski},\ and\ \citenamefont {Benson}}]{SMBH_Merger_Rate}%
  \BibitemOpen
\bibfield  {number} {  }\bibfield  {author} {\bibinfo {author} {\bibfnamefont {A.~L.}\ \bibnamefont {Erickcek}}, \bibinfo {author} {\bibfnamefont {M.}~\bibnamefont {Kamionkowski}},\ and\ \bibinfo {author} {\bibfnamefont {A.~J.}\ \bibnamefont {Benson}},\ }\bibfield  {title} {\bibinfo {title} {Supermassive black hole merger rates: uncertainties from halo merger theory},\ }\href {https://doi.org/10.1111/j.1365-2966.2006.10838.x} {\bibfield  {journal} {\bibinfo  {journal} {Monthly Notices of the Royal Astronomical Society}\ }\textbf {\bibinfo {volume} {371}},\ \bibinfo {pages} {1992} (\bibinfo {year} {2006})},\ \Eprint {https://arxiv.org/abs/https://academic.oup.com/mnras/article-pdf/371/4/1992/3638436/mnras0371-1992.pdf} {https://academic.oup.com/mnras/article-pdf/371/4/1992/3638436/mnras0371-1992.pdf} \BibitemShut {NoStop}%
\bibitem [{\citenamefont {Middleton}\ \emph {et~al.}(2021)\citenamefont {Middleton}, \citenamefont {Sesana}, \citenamefont {Chen}, \citenamefont {Vecchio}, \citenamefont {Del Pozzo},\ and\ \citenamefont {Rosado}}]{Middleton_2021}%
  \BibitemOpen
  \bibfield  {author} {\bibinfo {author} {\bibfnamefont {H.}~\bibnamefont {Middleton}}, \bibinfo {author} {\bibfnamefont {A.}~\bibnamefont {Sesana}}, \bibinfo {author} {\bibfnamefont {S.}~\bibnamefont {Chen}}, \bibinfo {author} {\bibfnamefont {A.}~\bibnamefont {Vecchio}}, \bibinfo {author} {\bibfnamefont {W.}~\bibnamefont {Del Pozzo}},\ and\ \bibinfo {author} {\bibfnamefont {P.~A.}\ \bibnamefont {Rosado}},\ }\bibfield  {title} {\bibinfo {title} {Massive black hole binary systems and the nanograv 12.5 yr results},\ }\href {https://doi.org/10.1093/mnrasl/slab008} {\bibfield  {journal} {\bibinfo  {journal} {Monthly Notices of the Royal Astronomical Society: Letters}\ }\textbf {\bibinfo {volume} {502}},\ \bibinfo {pages} {L99} (\bibinfo {year} {2021})},\ \Eprint {https://arxiv.org/abs/https://academic.oup.com/mnrasl/article-pdf/502/1/L99/54638697/slab008.pdf} {https://academic.oup.com/mnrasl/article-pdf/502/1/L99/54638697/slab008.pdf} \BibitemShut {NoStop}%
\bibitem [{\citenamefont {Banik}\ \emph {et~al.}(2018)\citenamefont {Banik}, \citenamefont {Tan},\ and\ \citenamefont {Monaco}}]{SMBHNumberDensity}%
  \BibitemOpen
  \bibfield  {author} {\bibinfo {author} {\bibfnamefont {N.}~\bibnamefont {Banik}}, \bibinfo {author} {\bibfnamefont {J.~C.}\ \bibnamefont {Tan}},\ and\ \bibinfo {author} {\bibfnamefont {P.}~\bibnamefont {Monaco}},\ }\bibfield  {title} {\bibinfo {title} {The formation of supermassive black holes from population iii.1 seeds. i. cosmic formation histories and clustering properties},\ }\href {https://doi.org/10.1093/mnras/sty3298} {\bibfield  {journal} {\bibinfo  {journal} {Monthly Notices of the Royal Astronomical Society}\ }\textbf {\bibinfo {volume} {483}},\ \bibinfo {pages} {3592} (\bibinfo {year} {2018})}\BibitemShut {NoStop}%
\bibitem [{\citenamefont {Abbott}\ \emph {et~al.}(2021)\citenamefont {Abbott} \emph {et~al.}}]{LIGO_Merger_Rate}%
  \BibitemOpen
  \bibfield  {author} {\bibinfo {author} {\bibfnamefont {R.}~\bibnamefont {Abbott}} \emph {et~al.},\ }\bibfield  {title} {\bibinfo {title} {Population properties of compact objects from the second ligo–virgo gravitational-wave transient catalog},\ }\href {https://doi.org/10.3847/2041-8213/abe949} {\bibfield  {journal} {\bibinfo  {journal} {The Astrophysical Journal Letters}\ }\textbf {\bibinfo {volume} {913}},\ \bibinfo {pages} {L7} (\bibinfo {year} {2021})}\BibitemShut {NoStop}%
\end{thebibliography}%

\end{document}